\documentclass[prd,amsfonts,onecolumn,preprint,nofootinbib,showpacs]{revtex4}

\usepackage{graphicx}


\begin{document}

\preprint{IPPP/08/82}
\preprint{DCPT/08/164}

\title{Cosmological scenario of stop NLSP with gravitino LSP \\ and
the cosmic lithium problem}

\author{Kazunori Kohri$^{1}$} 
\email{k.kohri@lancaster.ac.uk}

\author{Yudi Santoso$^{2}$}
\email{yudi.santoso@durham.ac.uk}

\affiliation{
$^ 1$ Physics Department, Lancaster University,~Lancaster~LA1~4YB,~UK \\
$^ 2$ Institute for Particle Physics Phenomenology,
Department of Physics, University of Durham, Durham DH1 3LE, UK
}

\begin{abstract}
The discrepancy on $^7$Li and $^6$Li abundances between the observational data and the standard Big Bang Nucleosynthesis theory prediction has been a nagging problem in astrophysics and cosmology, given the highly attractive and succesful Big Bang paradigm. One possible solution of this lithium problem is through hadronic decays of a massive metastable particle which alter the primordial element abundances. We explore this possibility using gravitino dark matter framework in which the next lightest supersymmetric particle (NLSP) is typically long-lived. We found that stop NLSP can provide an attractive solution to the lithium problem.
\end{abstract}

\maketitle

%%%%%%%%%%%%%%%%%%%%%%%%%%%%%%%%%%%%%%%%%%%%%%%%%%%%%%%%%%%%%%%%%%%%%%
\section{Introduction}
%%%%%%%%%%%%%%%%%%%%%%%%%%%%%%%%%%%%%%%%%%%%%%%%%%%%%%%%%%%%%%%%%%%%%%
\label{sec:introduction}

The identity of dark matter is one of the most important questions we
currently have in astrophysics and cosmology. There are many
theoretical candidates for dark matter particles proposed by many
beyond the standard model hypotheses in particle physics, and
gravitino is one of them within the supergravity
framework~\cite{Pagels:1981ke,Ellis:1983ew,Moroi:1993mb}. Due to the smallness of
the coupling between gravitino and matter in supergravity models,
$\sim 1/M_{\rm Pl}$, the next lightest supersymmetric particle (NLSP)
is typically longlived. This has direct implication on the light
element abundances if the NLSP decayed around or after the time of the
Big Bang Nucleosynthesis (BBN). Thus BBN provides a stringent
constraint on gravitino dark matter
scenarios~\cite{Ellis:2003dn,Feng:2004zu,Feng:2004mt,Cerdeno:2005eu,Steffen:2006hw,Kanzaki:2006hm,
Cyburt:2006uv,DiazCruz:2007fc,Kawasaki:2007xb,kawasaki:2008,
Ellis:2008as}. However, on the other hand it seems that the standard
BBN theory cannot explain all of the primordial light element
abundances from observations, namely the lithium problem described
below. This leads to an interesting possibility that some non-standard
processes were involved in the BBN. In this paper we study feasible
solution of the lithium problem by gravitino dark matter with stop
NLSP scenario.
 
There are many possibilities for the NLSP with the gravitino LSP, and
each has its own phenomenological signatures. The ones that have been
studied are neutralino
NLSP~\cite{Ellis:2003dn,Feng:2004mt,Cerdeno:2005eu,Steffen:2006hw,Cyburt:2006uv,kawasaki:2008}, stau
NLSP~\cite{Ellis:2003dn,Cerdeno:2005eu,Steffen:2006hw,Cyburt:2006uv,Kawasaki:2007xb,kawasaki:2008},
stop NLSP~\cite{DiazCruz:2007fc} and sneutrino
NLSP~\cite{Feng:2004zu,Kanzaki:2006hm,kawasaki:2008,Ellis:2008as}. We
do not discuss the details of those phenomenologies here, but
concentrate only on the BBN effects by looking at the NLSP decays,
lifetime and density before the decay. We found that stop NLSP is the
most interesting scenario in the sense that it provides possible
solution to the lithium problem. Therefore we focus on stop particle in this
paper.
 
It has been recognized that the theoretical prediction of $^{7}$Li
abundance in the standard BBN does not agree with the observational
results when we adopt the baryon-to-photon ratio of WMAP 5-year,
$\eta$ = $(6.225 \pm 0.170) \times
10^{-10}$~\cite{Dunkley:2008ie}. Using this ratio, the theoretical
value for $^{7}$Li is much larger than the observational one even if
we adopt a relatively high value of the observational abundance,
$\log_{10}(^{7}{\rm Li/H})_{\rm obs} = -9.36 \pm
0.06$~\cite{Melendez:2004ni}~\footnote{On the other hand in
Ref.~\cite{Bonifacio:2006au} a relatively low value of $^{7}$Li
abundance was reported, $\log_{10}(^{7}{\rm Li/H})_{\rm obs} = -9.90
\pm 0.06$. Obviously it would be more difficult and problematic to fit
this value.}. Quite recently Refs~\cite{Li7problem} reported that the
discrepancy gets worse if we adopt an updated theoretical calculation
for the reaction rate of $^{4}$He($^{3}$He,$\gamma$)$^{7}$Be.  As for
$^{6}$Li abundance, on the other hand, recent observation shows that
the theoretical value is much smaller than that of the observation,
($^{6}$Li/$^{7}$Li)$_{\rm obs}$ = 0.046 $\pm$
0.022~\cite{Asplund:2005yt}.  We collectively call these discrepancies
of $^{6}$Li and $^{7}$Li as the ``lithium problem''.

If there is a metastable massive particle which decays, producing
energetic standard model particles, around or after the BBN era, the
light element abundances might be altered through electromagnetic and
hadronic shower effects.  In the hadronic-decay scenario, we might be
able to solve the lithium problem because the hadron-emission could
possibly reduce $^{7}$Li~\cite{Jedamzik:2004er} and produce $^{6}$Li
simultaneously~\cite{Jedamziketal:2006,Cumberbatch:2007me} as will be
discussed later.  To solve the lithium problem, the NLSP abundance
times its net visible energy which fragments into hadrons $E_{\rm
had}$ should be in the range of $E_{\rm had}~n_{\rm NLSP}/s \simeq
10^{-14}$~GeV -- $10^{-13}$~GeV with their lifetime $10^{3}$ --
$10^{4}$
sec~\cite{Jedamzik:2004er,Jedamziketal:2006,Cumberbatch:2007me}.  It
is attractive that the stop abundance is naturally tuned to solve the
lithium problem when we consider a further (second) annihilation which
must occur just after the QCD phase transition as was pointed out
by~\cite{Kang:2006yd}.~\footnote{For another simultaneous solutions to
solve the lithium problem in astrophysics and cosmology, see also
Refs.~\cite{Korn:2006tv,Cayrel:2008hk}
and~\cite{Bird:2007ge,Jittoh:2008eq,Hisano:2008ti}, respectively.}

The outline of this paper is as follows. In section~2 we discuss supersymmetric models with stop NLSP, assuming gravitino as the LSP.  In section~3 we calculate the stop abundance before it eventually decays to gravitino. In section~4 we briefly review the Big Bang Nucleosynthesis theory and then presented our analysis with  stop NLSP. Finally, we conclude in section~5.  

%%%%%%%%%%%%%%%%%%%%%%%%%%%%%%%%%%%%%%%%%%%%%%%%%%%%%%%%%%%%%%%%%%%%%%
\section{Models with stop NLSP}
%%%%%%%%%%%%%%%%%%%%%%%%%%%%%%%%%%%%%%%%%%%%%%%%%%%%%%%%%%%%%%%%%%%%%%
\label{sec:stopmodel}

The lightest stop particle, $\widetilde{t}_1$, can be light by (little) seesaw mechanism between left and right stop, $\widetilde{t}_L$ and $\widetilde{t}_R$ respectively. Recall the stop mass matrix
\begin{equation}
{\cal M}_{\widetilde{t}}^2  = \left( \begin{array}{cc} M_{LL}^2 & M_{LR}^2 \\ M_{LR}^{2 \dag} & M_{RR}^2 \end{array} \right)
\end{equation}
where
\begin{eqnarray}
M_{LL}^2 &=&  M_{\widetilde{t}_L}^2 + m_t^2 + \frac{1}{6} (4 \, m_W^2 - m_Z^2) \cos 2 \beta \\
M_{RR}^2 &=&  M_{\widetilde{t}_R}^2 + m_t^2 + \frac{2}{3} m_Z^2 \cos 2 \beta \sin^2 \theta_W \\
M_{LR}^2 &=& - m_t (A_t + \mu \cot \beta)
\end{eqnarray}
The off diagonal terms are multiplied by the top-quark mass $m_t$ which is quite large. If $A_t$ is also large, it is possible to have $\widetilde{t}_1$ lighter than all the other MSSM sparticles, and therefore becomes the NLSP (with gravitino as the LSP). 

As we mentioned, in this gravitino LSP scenario, the stop NLSP has a
long lifetime.  In colliders, any metastable stop produced would
quickly hadronize either into some sbaryons or mesinos. The
super-hadrons then decay to the lightest sbaryon $\widetilde{t} ud$,
which is charged, or the lightest mesino, $\widetilde{t} \bar{u}$,
which is neutral~\cite{Gates:1999ei}~\footnote{For simplicity we will ignore the label `1'
to denote the lightest state for the rest of this paper.}. They would
then escape the calorimeter, and the charged ones could be detected by
the muon detector. This hadronization complicates the determination of
the detection rates of stop. Nevertheless, there has been analysis by
CDF at Tevatron which set the lower bound of (meta)stabe stop mass at
$\sim 250$~GeV~\cite{nachtman}.~\footnote{There are also studies 
for future detectabilities of a relatively long-lived stop in the Large Hadron Collider (LHC)~\cite{Allanach:2006fy} and the International Linear Collider (ILC)~\cite{Freitas:2007zr}. However, these studies assume neutralino LSP and the analyses were based on the stop decay to charm-quark plus neutralino, and therefore not applicable to our case here. 
}

Realization of stop NLSP scenario in some specific MSSM models was
studied in~\cite{DiazCruz:2007fc}.  However, it was found that in the Constrained MSSM
(CMSSM) it is not possible to have stop NLSP due to the Higgs mass and
the stop mass lower bound constraints. In the Non Universal Higgs
Masses (NUHM) model~\cite{Ellis:2002wv,Ellis:2002iu} we can still have a narrow allowed region with
stop NLSP. Nonetheless, stop NLSP is still possible if we forgo the
universality assumption for sfermion and gaugino masses. For example,
in~\cite{Martin:2007gf} it was shown that stop would be relatively light if the gluino mass at the GUT scale is lower than the bino and wino mass. In that paper the author still assumed neutralino be lighter than the stop. However, it would be easy to check that we can combine this non-universality with moderate value of trilinear coupling $A_t$ to get stop be the lightest. Our approach here is not to
look at specific model of supersymmetry. We will just assume that stop is the NLSP and treat the stop mass as a free parameter. 

\begin{figure}
    \begin{center}
        \includegraphics[width=130mm,clip,keepaspectratio]{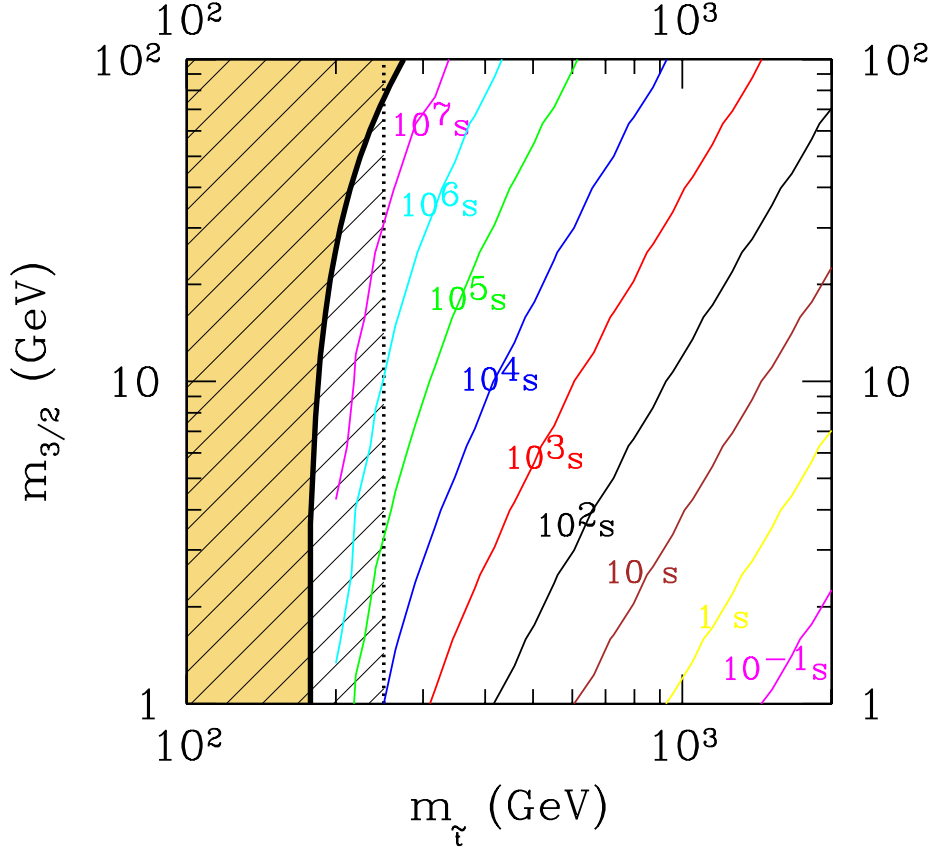}
\vskip -0.4in
    \caption{
        Stop lifetime in the plane of stop mass $m_{\widetilde{t}}$ and gravitino
        mass $m_{3/2} \equiv m_{\widetilde{G}}$. The shadowed region
        on the left with respect to the dotted line is excluded by
   the experimental lower bound on the stop mass 
      $m_{\widetilde{t}} \gtrsim 250$~GeV. 
        The left side of the thick line is not
        allowed by kinematic for 2-body
        decay, $\widetilde{t} \to t + \widetilde{G}$. }
        \label{fig:lifetime}
    \end{center}
\end{figure}

We then calculate the stop lifetime as a function of stop and gravitino masses. When the mass gap between stop and gravitino is larger than $m_t$ the dominant decay channel is the 2-body decay $\widetilde{t} \to \widetilde{G} + t$. The stop lifetime can then be determined as 
\begin{eqnarray}
\tau_{\widetilde{t}} &\simeq& 48 \pi M_{\rm P}^2 m_{\widetilde{G}}^2 m_{\widetilde{t}}^3 \left[ m_{\widetilde{t}}^2 - m_{ \widetilde{G}}^2 - m_t^2 + 4 \sin \theta_{\widetilde{t}} \cos \theta_{\widetilde{t}} m_t m_{\widetilde{G}} \right]^{-1} \nonumber \\
&& \times \left[ (m_{\widetilde{t}}^2 + m_{\widetilde{G}}^2 - m_t)^2 - 4 m_{\widetilde{t}}^2 m_{\widetilde{G}}^2 \right]^{-1} \left[(m_{\widetilde{t}}^2 + m_t^2 - m_{\widetilde{G}}^2 )^2 - 4 m_{\widetilde{t}}^2 m_t^2 \right]^{-1/2} \ ,
\end{eqnarray}
where $M_{\rm P}$ is the reduced Planck mass ($M_{\rm P}\equiv M_{\rm pl}/\sqrt{8\pi} \simeq 2.4 \times 10^{18}$~GeV) and $\theta_{\widetilde{t}}$ as the stop mixing angle. As shown in~\cite{DiazCruz:2007fc}, the dependence on $\theta_{\widetilde{t}}$ is small, hence we can fix $\theta_{\widetilde{t}}$ for our analysis (=1.3 radian). 
Our numerical results are shown in Fig.~\ref{fig:lifetime}, as contours in the $m_{\widetilde{t}}$ vs $m_{\widetilde{G}}$ plane. We do not need to calculate the 3-body decay rate since it is only important for lifetime longer than $\sim 10^8$~s.  

%%%%%%%%%%%%%%%%%%%%%%%%%%%%%%%%%%%%%%%%%%%%%%%%%%%%%%%%%%%%%%%%%%%%%%
%%%%%%%%%%%%%%%%%%%%%%%%%%%%%%%%%%%%%%%%%%%%%%%%%%%%%%%%%%%%%%%%%%%%%%

%%%%%%%%%%%%%%%%%%%%%%%%%%%%%%%%%%%%%%%%%%%%%%%%%%%%%%%%%%%%%%%%%%%%%%
\section{Stop abundance}
%%%%%%%%%%%%%%%%%%%%%%%%%%%%%%%%%%%%%%%%%%%%%%%%%%%%%%%%%%%%%%%%%%%%%%
\label{sec:abundance}

We can calculate the relic abundance of the thermally-produced stop
according to the standard method for calculating the  freeze-out
value~\cite{DiazCruz:2007fc}. Assuming that coannihilation effect is negligible and that there is no resonance, we can simplify the calculation by considering only annihilation channels through gluon exchange. See~\cite{Berger:2008ti} for detail  discussion on this approximation. In this case, the relic density calculation depends only on the stop mass. We get a relic density of $\Omega_{\widetilde{t}} \, h^2 = 10^{-4} - 10^{-2}$ for $m_{\widetilde{t}} = 10^2 - 10^3$~GeV, corresponding to $m_{\widetilde{t}} Y_{\widetilde{t}} = 10^{-13} - 10^{-11}$~GeV. However, in addition to the standard annihilation process, there is a further annihilation that reduces the number density of stop before it eventually decays~\cite{Kang:2006yd}.

After the QCD phase transition which occurs at  a temperature $T =
T_{\rm QCD} \approx 150$~MeV, all of the strong interacting particles
are confined into hadrons. The strength of the strong interaction is
then determined  by the effective theory of hadron physics. According
to the heavy-quark effective
theory~\cite{Wise:1994vh}, the length scale for the
interactions of heavy-quark hadrons is $R_{\rm had} \sim$~1 --
10~GeV$^{-1}$~\cite{Kang:2006yd}, and we can parameterize the length scale  as
$R_{\rm had} \equiv f_{\sigma} /m_{\pi}$  with $m_{\pi}$ is the pion
mass ( $\simeq 140$~MeV) and $f_{\sigma}$ is a numerical parameter
ranging from about 0.1 to 1.

The relic stop and antistop would also form hadrons, called
super-hadrons or shadrons, which then undergo further
annihilation via bound states. The annihilation rate among these shadrons can be
written as
\begin{eqnarray}
    \label{eq:GammaAnn}
    \Gamma_{\rm ann} =  \langle \sigma v \rangle \, n_{\widetilde{t}} \ ,
\end{eqnarray}
where $\sigma$ is the annihilation cross section, $v$ is the
relative velocity, and the bracket means thermal average. Here we
assume that all the stop shadrons have decayed into the lightest one which is a neutral mesino, $\widetilde{T}^0 \equiv \widetilde{t} \bar{u}$, and similarly $\widetilde{T}^{0 \ast} \equiv \widetilde{t}^\ast u$ for the anti-stops. With this assumption, the number density of $\widetilde{T}^0$ is equal to that of the stop
$n_{\widetilde{t}}$~\cite{DiazCruz:2007fc}. We also assume that the number density of stop is equal to that of anti-stop. Here $\sigma$ and $v$ can be
simply  expressed by $\sigma \sim R_{\rm had}^{2}$ and  $v \simeq
\sqrt{3T/m_{\widetilde{t}}}$ \, .

The timescale of the annihilation is compared with the Hubble
expansion rate,
\begin{eqnarray}
    \label{eq:3Hubble}
    \Gamma_{\rm ann} = 3 H,
\end{eqnarray}
where $H = \sqrt{\rho}/(\sqrt{3} M_{\rm P})$ with $ M_{\rm P}$ is the
reduced Planck mass, and $\rho$ is the total energy density, $\rho =
\pi^{2}/30 g_{*} T^{4}$. We adopt effective degrees of freedom $g_{*}
= 17.25$ (or 10.75), assuming that the QCD phase transition occurs at
$T = T_{\rm QCD} \sim 150$ MeV (or $\sim 100$ MeV).  We get the final
abundance of the stop after the second annihilation to
be
\begin{eqnarray}
    \label{eq:stopyield}
    m_{\widetilde{t}} Y_{\widetilde{t}} 
    = 0.87 \times 10^{-14} \, {\rm GeV} \left(
    \frac{f_{\sigma}}{0.1} \right)^{-2} \left( \frac{g_{*}}{17.25}
    \right)^{-1/2} \left( \frac{T_{\rm QCD}}{150 \, {\rm MeV}}
    \right)^{-3/2} \left( \frac{m_{\widetilde{t}}}{10^{2} \, {\rm
    GeV}} \right)^{3/2},
\end{eqnarray}
where the yield variable of the stop is defined by $Y_{\widetilde{t}}
\equiv n_{\widetilde{t}}/s$ with $s$ is the entropy density (=$\frac43
\rho /T$).  From the temperature dependence in
Eq.(\ref{eq:stopyield}), it can be seen that $Y_{\widetilde{t}}$ is
lower for a higher temperature after the QCD phase transition, which
means that the number density of the stop must immediately be
frozen-out just after their hadronization and annihilation at $T =
T_{\rm QCD}$. We see that the stop abundance can naturally be cast
into the attractive range of $10^{-14} \, {\rm GeV} \lesssim
B_{h}m_{\widetilde{t}} Y_{\widetilde{t}} \lesssim 10^{-13} \, {\rm
GeV}$~\footnote{To be more precise it is
$B_{h}(m_{\widetilde{t}}-m_{\widetilde{G}}) Y_{\widetilde{t}}$ that we
need to look at. However, the stop mass is much larger than the
gravitino mass in the interesting region.} for a several hundred GeV
stop mass with a reasonable hadronic branching ratio $B_{h} \sim {\cal
O}(1)$.  Note that the above final abundance is lower than the stop
abundance after the standard freeze-out, which occurred at around $T
\sim m_{\widetilde{t}}/30$~\cite{DiazCruz:2007fc}.  Moreover, the
decay of the thermally-produced stop would have only negligible
contribution to the dark matter relic density $\Omega_{\rm DM} h^{2}
\simeq 0.1$~\cite{Dunkley:2008ie}.

%%%%%%%%%%%%%%%%%%%%%%%%%%%%%%%%%%%%%%%%%%%%%%%%%%%%%%%%%%%%%%%%%%%%%%
\section{Big-bang nucleosynthesis and stop NLSP}
%%%%%%%%%%%%%%%%%%%%%%%%%%%%%%%%%%%%%%%%%%%%%%%%%%%%%%%%%%%%%%%%%%%%%%
\label{sec:BBN}

We consider the 2-body decay process of stop into a gravitino and a
top quark. The emitted top quark then immediately fragments into
hadrons and produce lots of high-energy protons and neutrons. Those
emitted particles may modify the abundances of light elements such as
D, T, $^{3}$He, $^{4}$He, $^{6}$Li, $^{7}$Li and $^{7}$Be after/during
the BBN epoch~\cite{Reno:1987qw,Dimopoulos:1987fz,Kawasaki:2004yh,
Jedamzik:2006xz}.

The high-energy hadrons scatter off the background protons and
$^{4}$He, and induce hadronic shower. These processes produce
energetic neutron, D, T and $^{3}$He. These non-thermally produced
neutron and T (or $^{3}$He) then scatter off the background protons
and $^{4}$He. It is followed by $n$-$p$ (T-$^{4}$He) reactions,
synthesizing D ($^{6}$Li). In addition, the nonthermally-produced
neutron also induces sequential reactions to reduce the $^{7}$Be
abundance through a set of processes $^{7}$Be($n$,
p)$^{7}$Li(p,$^{4}$He)$^{4}$He, reducing the primordial $^{7}$Be
abundance which means reducing the primordial $^{7}$Li abundance as
well through the electron capture at a later time. This mechanism in
non-standard BBN scenarios have been studied
by~\cite{Jedamzik:2004er,Jedamziketal:2006,Cumberbatch:2007me} in
detail.  Because we are interested in a relatively short lifetime of
stop $\tau_{\widetilde{t}} \lesssim 10^{7} s$ with its high branching
ratio into hadrons ($B_{h} \sim 1$), the photodissociation processes
induced by the emitted high-energy charged particles and photons are
not important in this study~\cite{Kawasaki:2004yh,Jedamzik:2006xz}.

We would have to check  whether the abundances of the other light elements 
also agree with the observational data. In particular, it would be a crucial problem if copious deuteriums are produced in this scenario.  For the observational
deuterium abundance, we adopt (D/H)$_{\rm obs} = (2.82 \pm 0.26)
\times 10^{-5}$~\cite{O'Meara:2006mj}. Because the stop
abundance before its decay is very small as was discussed in the previous section, the stop decay does not significantly affect the $^{4}$He mass fraction and the $^{3}$He to D ratio. 

\begin{figure}
    \begin{center}
        \includegraphics[width=160mm,clip,keepaspectratio]{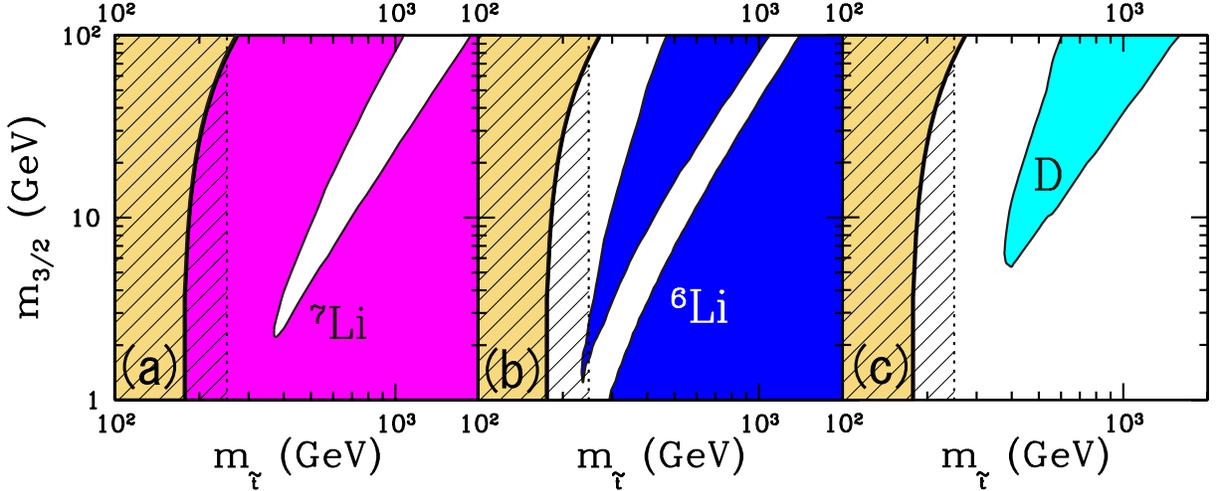}
        \caption{
        Allowed regions in the plane of the stop mass and the gravitino
        mass in case of $R_{\rm had} = 0.1/m_{\pi}$.  Each white
        region is observationally allowed for (a) $^{7}$Li, (b)
        $^{6}$Li and (c) D, respectively. The shadowed region
        which is left with respect to the dotted line is excluded by
        the experimental lower bound $m_{\widetilde{t}} \gtrsim 250$
        GeV. The left region with respect
        to the thick line is not allowed kinematically when we
        consider only the two-bodies decay, $\widetilde{t} \to t +
        \widetilde{G}$. }
        \label{fig:mstopm32}
    \end{center}
\end{figure}

In Fig.~\ref{fig:mstopm32} we plot allowed regions in the plane of the
stop mass and the gravitino mass,  assuming $f_{\sigma} = 0.1$.
Each white region is observationally allowed for (a) $^{7}$Li, (b)
$^{6}$Li and (c) D, respectively. The left region with respect to the
thick line is not kinematically allowed for the 2-body decay mode, $\widetilde{t} \to t + \widetilde{G}$, and has stop lifetime longer than $\sim 10^9$~s which we do not consider here. Note that even if we include this region our results would not be changed. Furthermore, there is experimental constraint which excludes $m_{\widetilde{t}} \lesssim 250$~GeV. 

%%%%%%%%%%%%%%%%%%%%%%%%%%%%%%%%%%%%%%%%%%%%%%%%%%%%%%%%%%%%%%%%%%%%%%
%%%%%%%%%%%%%%%%%%%%%%%%%%%%%%%%%%%%%%%%%%%%%%%%%%%%%%%%%%%%%%%%%%%%%%

We can understand the exclusion/allowed regions in Fig.~\ref{fig:mstopm32} by looking at the yield $B_{h} m_{\widetilde{t}} Y_{\widetilde{t}}$ and the lifetime $\tau_{\widetilde{t}}$. The analysis results in terms of these two variables is shown in Fig.~1 of Ref.~\cite{Cumberbatch:2007me}. 
From Eq.(8) we found that the yield of the stop grows as $m_{\widetilde{t}}^{3/2}$ as the stop mass increases. For small stop mass the effect on $^7$Li is too small, while for large mass the effect is too large. It is just right in between. In addition the effect also depends on the stop lifetime which can be seen from Fig.~\ref{fig:lifetime}. While for $^6$Li, as shown in Fig.~1 of~\cite{Cumberbatch:2007me}, it prefers small yield with relatively large lifetime, or a certain range of lifetime almost independent of the yield. As for D, it is allowed if the yield is small or the lifetime is small.

\begin{figure}
    \begin{center}
        \includegraphics[width=130mm,clip,keepaspectratio]{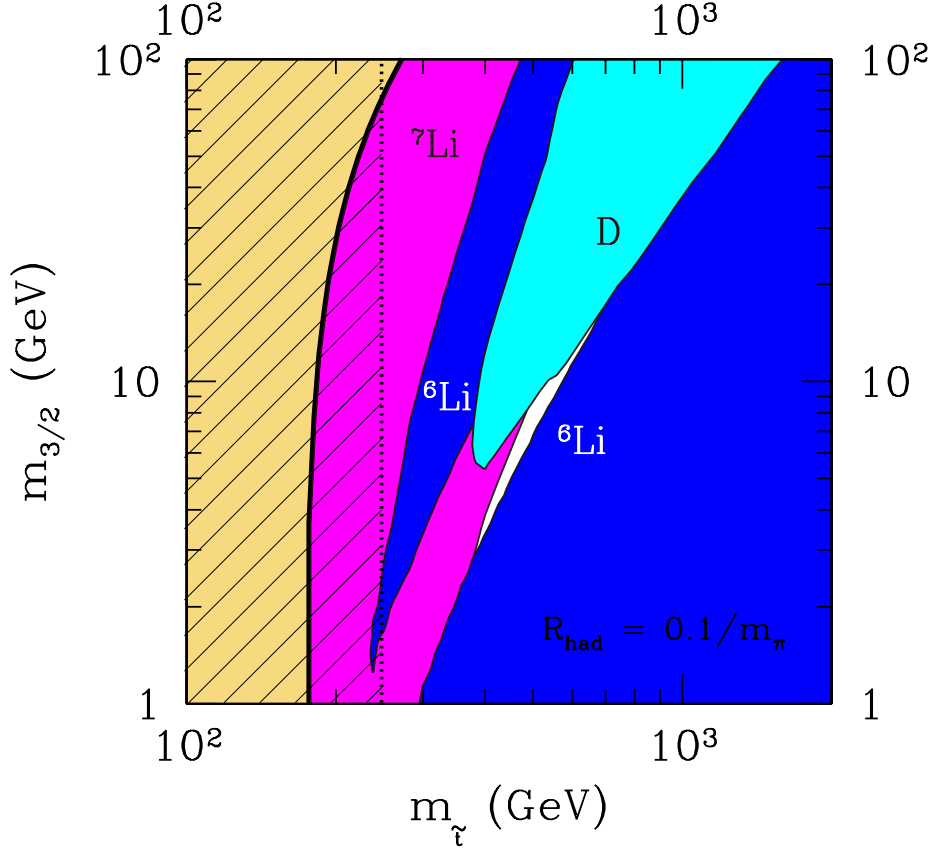}
\vskip -0.4in
        \caption{
        Combined allowed region in the plane of the stop mass and the
        gravitino mass by using the constrains in
        Fig~\ref{fig:mstopm32} (a), (b), and (c).  The name of each
        element is written in the respective exclusion regions. There is
        an allowed region at around $m_{\widetilde{t}} = 400 - 600$~GeV
        and $m_{3/2} = 2 - 10$~GeV.}
        \label{fig:comb}
    \end{center}
\end{figure}

By combining all three constraints shown in Fig.~\ref{fig:mstopm32}
(a),(b),(c), we get the favoured region at around
$m_{\widetilde{t}} = 400 - 600$~GeV and $m_{3/2} = 2 - 10$~GeV which is
clearly shown in Fig.~\ref{fig:comb}.

%%%%%%%%%%%%%%%%%%%%%%%%%%%%%%%%%%%%%%%%%%%%%%%%%%%%%%%%%%%%%%%%%%%%%%
%%%%%%%%%%%%%%%%%%%%%%%%%%%%%%%%%%%%%%%%%%%%%%%%%%%%%%%%%%%%%%%%%%%%%%

%%%%%%%%%%%%%%%%%%%%%%%%%%%%%%%%%%%%%%%%%%%%%%%%%%%%%%%%%%%%%%%%%%%%%%
\section{Conclusions}
%%%%%%%%%%%%%%%%%%%%%%%%%%%%%%%%%%%%%%%%%%%%%%%%%%%%%%%%%%%%%%%%%%%%%%
We have studied the decays of stop NLSP to gravitino LSP and the effects on BBN. We found some range of stop and gravitino masses where all light element abundances agree with the observational data, including $^7$Li and $^6$Li which are problematic in the standard BBN. It is, therefore, very interesting to test this hypothesis further. The gravitino itself would be practically undetectable, aside from its gravitational effect, due to its very weak interaction. Direct detection of a dark matter particle can therefore exclude our model~\footnote{It might still be allowed if gravitino is not the only dark matter, and the other dark matter particle is not neutralino, i.e. decoupled from gravitino.}. We can only hope that future collider experiments, such as the upcoming LHC, would be able to provide some evidence on the physics beyond the standard model, and advance our understanding of the Early Universe. 

%%%%%%%%%%%%%%%%%%%%%%%%%%%%%%%%%%%%%%%%%%%%%%%%%%%%%%%%%%%%%%%%%%%%%%
\section{Acknowledgements}
%%%%%%%%%%%%%%%%%%%%%%%%%%%%%%%%%%%%%%%%%%%%%%%%%%%%%%%%%%%%%%%%%%%%%%
The work of KK  is partially supported by PPARC grant PP/D000394/1 and
by EU grants MRTN-CT-2004-503369 and MRTN-CT-2006-035863, the European
Union through the Marie Curie Research and Training Network
"UniverseNet". YS would like to thank Keith Olive for useful discussions.

%%%%%%%%%%%%%%%%%%%%%%%%%%%%%%%%%%%%%%%%%%%%%%%%%%%%%%%%%%%%%%%%%%%%%%

%%%%%%%%%%%%%%%%%%%%%%%%%%%%%%%%%%%%%%%%%%%%%%%%%%%%%%%%%%%%%%%%%%%%%%

\end{document}